\documentclass[twoside,english,aps,prl,superscriptaddress,twocolumn]{revtex4-1}
\usepackage[T1]{fontenc}
\usepackage[latin9]{inputenc}
\usepackage{xcolor}
\usepackage{verbatim}
\usepackage{amsmath}
\usepackage{amssymb}
\usepackage{graphicx}

\makeatletter

\@ifundefined{textcolor}{}
{%
 \definecolor{BLACK}{gray}{0}
 \definecolor{WHITE}{gray}{1}
 \definecolor{RED}{rgb}{1,0,0}
 \definecolor{GREEN}{rgb}{0,1,0}
 \definecolor{BLUE}{rgb}{0,0,1}
 \definecolor{CYAN}{cmyk}{1,0,0,0}
 \definecolor{MAGENTA}{cmyk}{0,1,0,0}
 \definecolor{YELLOW}{cmyk}{0,0,1,0}
}

\usepackage{palatino}

\usepackage[hypertexnames=false]{hyperref}
\hypersetup{colorlinks=true,citecolor=blue,linkcolor=red}

\AtBeginDocument{
\renewcommand{\ref}[1]{\autoref{#1}}

}
\makeatother

\usepackage{babel}
\begin{document}

\title{Imaging the Nanoscale Band Structure of Topological Sb \medskip{}
}

\author{Anjan Soumyanarayanan}

\email{anjan@physics.harvard.edu}

\affiliation{Department of Physics, Harvard University, Cambridge, MA 02138, USA}

\affiliation{Department of Physics, Massachusetts Institute of Technology, Cambridge,
MA 02139, USA}

\author{Michael M. Yee}

\affiliation{Department of Physics, Harvard University, Cambridge, MA 02138, USA}

\author{Yang He}

\affiliation{Department of Physics, Harvard University, Cambridge, MA 02138, USA}

\author{Hsin Lin}

\thanks{Present Address: Graphene Research Centre and Department of Physics,
National University of Singapore, Singapore 117542.}

\affiliation{Department of Physics, Northeastern University, Boston, MA 02115,
USA}

\author{Dillon R. Gardner}

\affiliation{Department of Physics, Massachusetts Institute of Technology, Cambridge,
MA 02139, USA}

\author{Arun Bansil}

\affiliation{Department of Physics, Northeastern University, Boston, MA 02115,
USA}

\author{Young S. Lee}

\affiliation{Department of Physics, Massachusetts Institute of Technology, Cambridge,
MA 02139, USA}

\author{Jennifer E. Hoffman}

\email{jhoffman@physics.harvard.edu}

\affiliation{Department of Physics, Harvard University, Cambridge, MA 02138, USA}
\begin{abstract}
\textcolor{black}{Many promising building blocks of future electronic
technology -- including non-stoichiometric compounds, strongly correlated
oxides, and strained or patterned films -- are inhomogeneous on the
nanometer length scale. }Exploiting the inhomogeneity of such materials
to design next-generation nanodevices requires a band structure probe
with nanoscale spatial resolution. To address this demand, we report
the first simultaneous observation and quantitative reconciliation
of two candidate probes -- Landau level spectroscopy and quasiparticle
interference imaging -- which we employ here to reconstruct the multi-component
surface state band structure of the topological semimetal antimony
(Sb). We thus establish the technique of band structure tunneling
microscopy (BSTM), whose unique advantages include nanoscale access
to non-rigid band structure deformation, empty state dispersion, and
magnetic field dependent states. We use BSTM to elucidate the relationship
between bulk conductivity and surface state robustness in topological
materials, and to quantify essential metrics for spintronics applications.
\end{abstract}
\maketitle

\section{Introduction\label{sec:intro}}

\textcolor{black}{Driven by novel components and fabrication techniques
for modern electronic devices\cite{Ando2013,Novoselov2005b}, it has
become imperative to develop a nanoscale understanding of the electronic
band structure \textendash{} the relationship between the electronic
energy and momentum -- within a wide variety of materials. T}he scanning
tunneling microscope (STM), best known for its atomic resolution imaging
capability, can also provide momentum-resolved ($k$-space) spectroscopic
information through two phenomena -- Landau quantization and quasiparticle
interference (QPI). First, the application of a magnetic field $B$
can quantize the electronic density of states (DOS) into Landau levels
(LLs), resulting in oscillations in the STM conductance ($dI/dV$)
spectra\cite{Morgenstern2000}. The LL dispersion can be mapped onto
the quasiparticle band structure in the semiclassical limit\cite{Miller2009a,Hanaguri2010}.
Increased disorder limits the formation of LLs, but enables the second
technique -- QPI imaging\cite{Crommie1993a}. Interference between
the initial and final wavevectors, $\vec{k}_{{\rm i}}$ and $\vec{k}_{{\rm f}}$,
of elastically scattered quasiparticles of energy $\varepsilon$,
can produce a standing wave pattern with wavevector $\vec{q}=\vec{k}_{{\rm f}}-\vec{k}_{{\rm i}}$
in the $dI/dV$ map at energy $\varepsilon=eV$, allowing the inversion
of $q(\varepsilon)$ to find $k(\varepsilon)$.

The nanoscale spatial resolution, temperature-limited energy resolution,
access to filled and empty states, and utility in magnetic field offered
by STM measurements of LLs and QPI make them ideal complements to
angle-resolved photoemission spectroscopy (ARPES) as band structure
probes. In fact, LL spectroscopy and QPI imaging have been cornerstone
techniques for over a decade, used to investigate gap symmetry in
superconductors\cite{Hoffman2002a,Hanaguri2009,Allan2013a}, backscattering
in topological materials\cite{Roushan2009,Zhang2009a,Okada2011},
pseudospin protection in graphene\cite{Rutter2007a,Zhang2009d} and
chemical potential fluctuations in a range of materials\cite{Wise2009,Zhang2009d,Beidenkopf2011,Okada2012c}.
Despite their tremendous promise,\emph{ }LLs and QPI \textbf{\emph{have
}}\textbf{\textit{never been simultaneously observed -- over the same
spatial area and energy range -- in any material}}; therefore the 
equivalence of these one- and two-particle phenomena has yet to be 
established. In fact, independent use of these techniques on graphene
have reported a $40\%$ discrepancy in Fermi velocity\cite{Luican2011a,Zhang2009d}.
Such discrepancies have been attributed to collective modes\cite{Zhang2009d},
variations in carrier density\cite{Chae2012a}, or tip-induced electric
fields\cite{Cheng2010} -- but the two techniques have never been
quantitatively reconciled. This problem undermines the widespread
use of LL spectroscopy and QPI imaging techniques.

Elemental Sb, of high current interest due to its nontrivial topology
and intriguing potential for spintronic devic\textcolor{black}{es,
provides an ideal platform to address this issue. Its n}egative band
gap guarantees sufficient bulk carrier density to screen chemical
potential fluctuations\cite{Roushan2009,Beidenkopf2011} and tip-induced
electric fields\cite{Cheng2010}, while its topological nature requires
the existence of robust surface states (SSs)\cite{Fu2007,Hsieh2009c,Chen2009,Hasan2010,Qi2011,Moore2010},
where LL and QPI phenomena may be observed. The topological surface
states derive from two spin-split parabolas which form inner and outer
Dirac cones connecting the valence and conduction bands (\ref{fig:bandstruct_topospectrum}a).
The surface states can be described by a five-parameter phenomenological
$k\cdot p$ Hamiltonian\cite{Fu2009}

\begin{equation}
H(k)=\varepsilon_{D}+\frac{k^{2}}{2m^{*}}+v_{0}(1+\alpha k^{2})(k_{x}\sigma_{y}-k_{y}\sigma_{x})+\frac{\lambda}{2}(k_{+}^{3}+k_{-}^{3})\sigma_{z}\label{eq:KPHamiltonian}
\end{equation}

\noindent where $\varepsilon_{D}$ is the Dirac point energy, $m^{*}$
is the effective mass, $\alpha$ \textcolor{black}{and $\lambda$
control the shapes of the two Dirac cones, and $v_{0}$ is the Rashba
parameter corresponding to the magnitude of spin-orbit coupling. This
five-parameter dispersion can serve as a key test case for comparing
the LL and QPI phenomena, while $v_{0}$ in particular is an essential
utility metric for spintronics devices.}

Here we report the simultaneous observation of LLs and QPI over a
300~meV energy range in Sb. We quantitatively reconcile these techniques
and\textbf{ }use them in concert to reconstruct the multi-component
surface state band structure, thus establishing the technique of \textbf{band
structure tunneling microscopy (BSTM)}. We demonstrate the nanoscale
spatial sensitivity of BSTM and use it to quantify several metrics
of Sb relevant to spintronics applications. More generally, we clarify
the relationship of topological surface states to proximate bulk bands,
thereby directing the wider exploration of technologically useful
topological materials.

\section{Results\label{sec:results}}

\begin{figure*}
\includegraphics[width=6.25in]{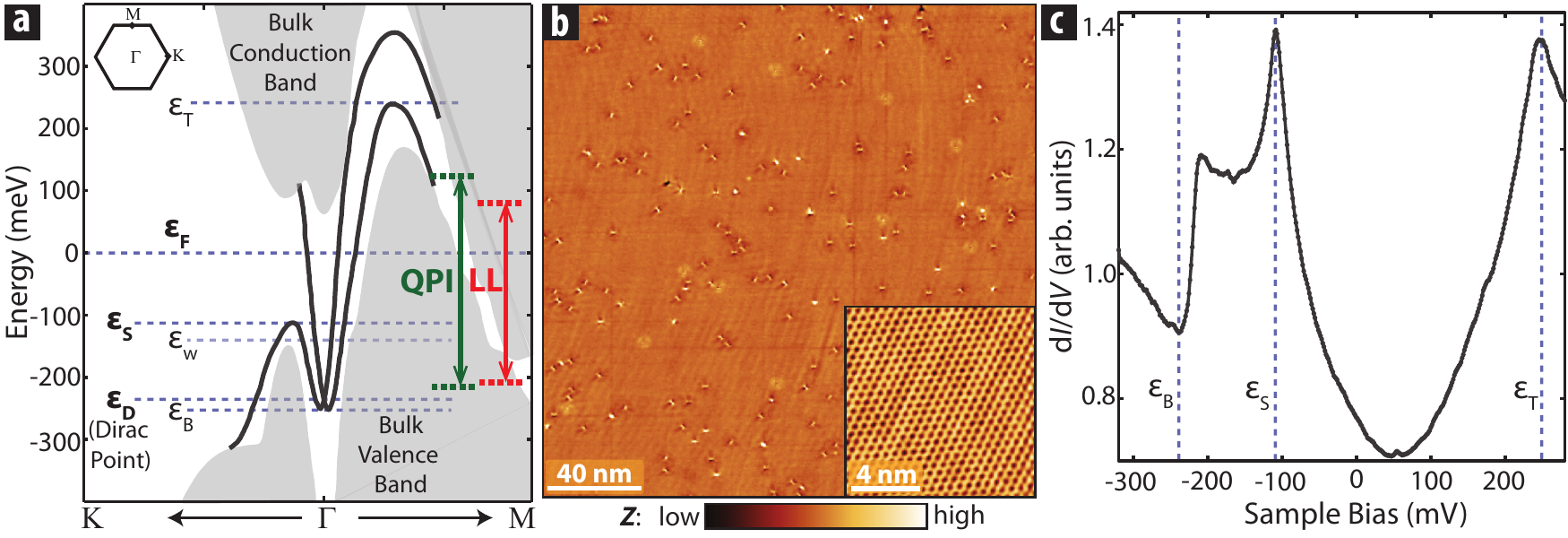}

\caption{\textbf{STM Topography and Band Structure.} \textbf{(a)} Schematic
band structure (from density functional theory) of the semimetal Sb(111)
with topological SSs (dark grey) connecting the bulk valence and conduction
bands (light grey). The spectral range of the observed LLs and QPI
is indicated. \textbf{(b)} STM topograph of Sb(111) showing an atomically
flat surface (sample bias, $V_{{\rm 0}}=+200\,{\rm mV}$; junction
resistance, $R_{{\rm J}}=10\:{\rm G}\Omega$). Inset shows the atomically
resolved hexagonal lattice ($V_{{\rm 0}}=+200\,{\rm mV}$; $R_{{\rm J}}=125\:{\rm M}\Omega$).
\textbf{(c)} Typical $dI/dV$ spectrum on Sb(111), with kinks at $\varepsilon_{{\rm B}}$,
$\varepsilon_{{\rm S}}$, and $\varepsilon_{{\rm T}}$, corresponding
to extremal features in the surface state band structure shown in
(a) ($V_{{\rm 0}}=+300\,{\rm mV}$; $R_{{\rm J}}=500\:{\rm M}\Omega$;
$V_{{\rm rms}}=3\,{\rm mV}$). \label{fig:bandstruct_topospectrum}}
\end{figure*}

Topographic STM images of the cleaved (111) surface of Sb (Supp. Info.
I) show large atomically flat regions (\ref{fig:bandstruct_topospectrum}b),
free from chemical potential fluctuations except in the immediate
vicinity of sparse single atom surface defects and step edges. The
$dI/dV$ spectrum (\ref{fig:bandstruct_topospectrum}c), proportional
to the local DOS, is dominated by cusp-like features associated with
extrema ($\varepsilon_{{\rm B}}$, $\varepsilon_{{\rm T}}$) and a
saddle point ($\varepsilon_{{\rm S}}$) in the SS band structure (\ref{fig:bandstruct_topospectrum}a).
The Dirac point is not directly visible due its spectral coincidence
with other SSs and bulk bands, however, the conducting bulk confers
the aforementioned benefits for momentum-resolved spectroscopic studies.

\begin{figure*}
\includegraphics[width=4.8in]{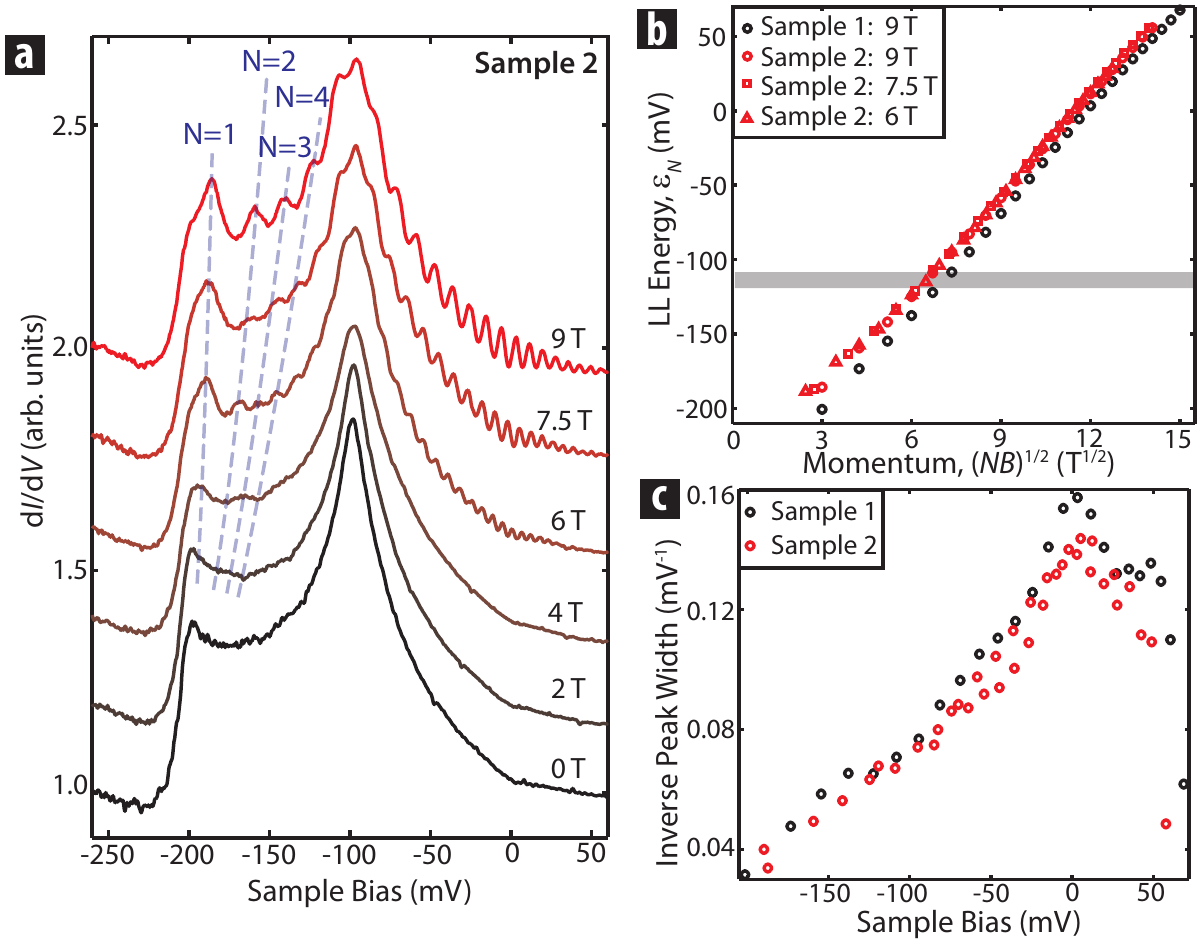}

\caption{\textbf{Landau Quantization of Surface States.} \textbf{(a)} Representative
$dI/dV$ spectra at several values of the magnetic field $B$, vertically
offset for clarity, showing the emergence of LL oscillations. Dashed
blue lines are guides to the eye following the first four LLs. Setpoint
Parameters: $V_{{\rm 0}}=+100$~mV; $R_{{\rm J}}=0.1-0.2\:{\rm G}\Omega$;
$V_{{\rm rms}}=0.4$~mV. \textbf{(b)} Dispersion of LL energies $\varepsilon_{N}$
vs. momentum $\sqrt{NB}$, based on the semiclassical approximation
for Dirac fermions (\ref{eq:ll_diracform}). Grey band shows the crossover
energy below which the LLs are described by the Rashba formula (\ref{eq:ll_rashbaform},
Supp Info II).\textbf{ }Between samples, the chemical potential and
dispersion vary by $\sim15$~mV and $\sim5\%$ respectively. \textbf{(c)}
The LL inverse peak widths, measured using Lorentzian fits (Supp Info
II), showing quasiparticle lifetime broadening away from $\varepsilon_{{\rm F}}$
(data acquired at 2.2~K). \label{fig:lls}}
\end{figure*}

In applied magnetic field $B$ above 4~T, Landau quantization causes
conductance oscillations to appear in the $dI/dV$ spectrum (\ref{fig:lls}a-b).
We assign empirical indices starting with $N=1$ to all such $B$-dependent
peaks (\ref{fig:lls}b, Supp. Info. II). We observe a remarkable 27~LLs
\textendash{} more than reported on any other topological material\cite{Hanaguri2010,Cheng2010,Okada2011,Jiang2012a}--
despite the presence of bulk bands throughout this energy range (\ref{fig:bandstruct_topospectrum}a).
The LL peaks are sharpest around the Fermi energy, $\varepsilon_{{\rm F}}$
(\ref{fig:lls}c), evincing monotonic quasiparticle lifetime broadening
away from $\varepsilon_{{\rm F}}$, in contrast to other topological
materials where collective modes complicate the picture\cite{Hanaguri2010,Okada2012b}.
The measured lifetime at $\varepsilon_{{\rm F}}$ corresponds to a
long elastic mean free path, $l_{{\rm f}}\sim65$~nm.

\textcolor{black}{}%
\textcolor{black}{We use the LLs, which correspond to closed contours
of constant energy (CCEs) in momentum space, to obtain part of the
SS dispersion on Sb in two energy regimes. First, we note that LLs
in other topological materials have been interpreted in the Dirac
fermion picture\cite{Hanaguri2010,Cheng2010,Okada2011,Jiang2012a},
with the energy of the $n^{{\rm th}}$ LL, $\varepsilon_{n}$, given
by }

\begin{equation}
\varepsilon_{n}(B)=\varepsilon_{{\rm D}}+v_{D\,}\sqrt{2e\hbar nB}\label{eq:ll_diracform}
\end{equation}

\noindent \textcolor{black}{where the Fermi velocity $v_{{\rm D}}$
is a constant over the energy range of interest. For }Dirac fermions,
the semiclassical Bohr-Sommerfeld quantization relation gives the
momentum space radius for the $n^{{\rm th}}$ LL orbit, $q_{n}=\sqrt{(2e/\hbar)\, nB}$~\cite{Hanaguri2010}.
\ref{fig:lls}b shows the LL peak energies, $\varepsilon_{N}$, plotted
against the empirical LL momentum, $q_{N}=\sqrt{NB}$. For energies
$\varepsilon>\varepsilon_{{\rm S}}$, the dispersions obtained at
different magnetic fields collapse on to a single curve \textendash{}
validating the Dirac fermion semiclassical approximation with $n=N$,
and demonstrating that the Landau quantization arises from a single
Dirac cone in this energy range. From \ref{fig:bandstruct_topospectrum}a,
we conclude that the LL wavevector $q_{N}$ corresponds to the radius
of the inner cone, and independently gives its velocity, $v_{{\rm LL}}\equiv v_{D}=4.20$~eV·Å
($6.34\times10^{5}$~${\rm m/{\rm s}}$).\textcolor{black}{{} Second,
for energies $\varepsilon<\varepsilon_{{\rm {\rm S}}}$, the presence
of two spin-split cones requires a different interpretation of the
LLs based on the Rashba picture\cite{Schliemann2003,Zarea2005}, where
$\varepsilon_{n}$ is given by:}

\noindent \textcolor{black}{
\begin{equation}
\varepsilon_{n}(B)=\begin{cases}
\varepsilon_{0}+\frac{1}{2}(\hbar\omega_{{\rm c}}+g\mu_{{\rm B}}B), & n=0\\
\varepsilon_{{\rm 0}}+\hbar\omega_{{\rm {\rm c}}}n\pm\sqrt{\delta^{2}/4+(2m^{*}\, v_{0}^{2})\cdot n\hbar\omega_{{\rm c}}}, & n>0
\end{cases}\label{eq:ll_rashbaform}
\end{equation}
Here, }$\varepsilon_{0}$ is the band offset,\textcolor{black}{{} $\omega_{{\rm c}}=eB/m^{*}$
is the cyclotron frequency, $\delta=(1-1/2\, gm^{*})\hbar\omega_{{\rm c}}$,
and $g$ is the electron g-factor\cite{Schliemann2003,Zarea2005}.
Our measured LLs correspond to the positive sign in \ref{eq:ll_rashbaform},
which derives predominantly from the inner Rashba cone, while the
LLs of the outer cone, with their tighter energy spacing, are not
visible due to broadening effects. Reconstructing both band structure
branches in this low energy ($\varepsilon<\varepsilon_{{\rm {\rm S}}}$)
regime therefore requires the combined use of the Rashba LLs and the
other momentum-resolved technique -- QPI.}

\noindent

\begin{figure*}
\includegraphics[width=4.6in]{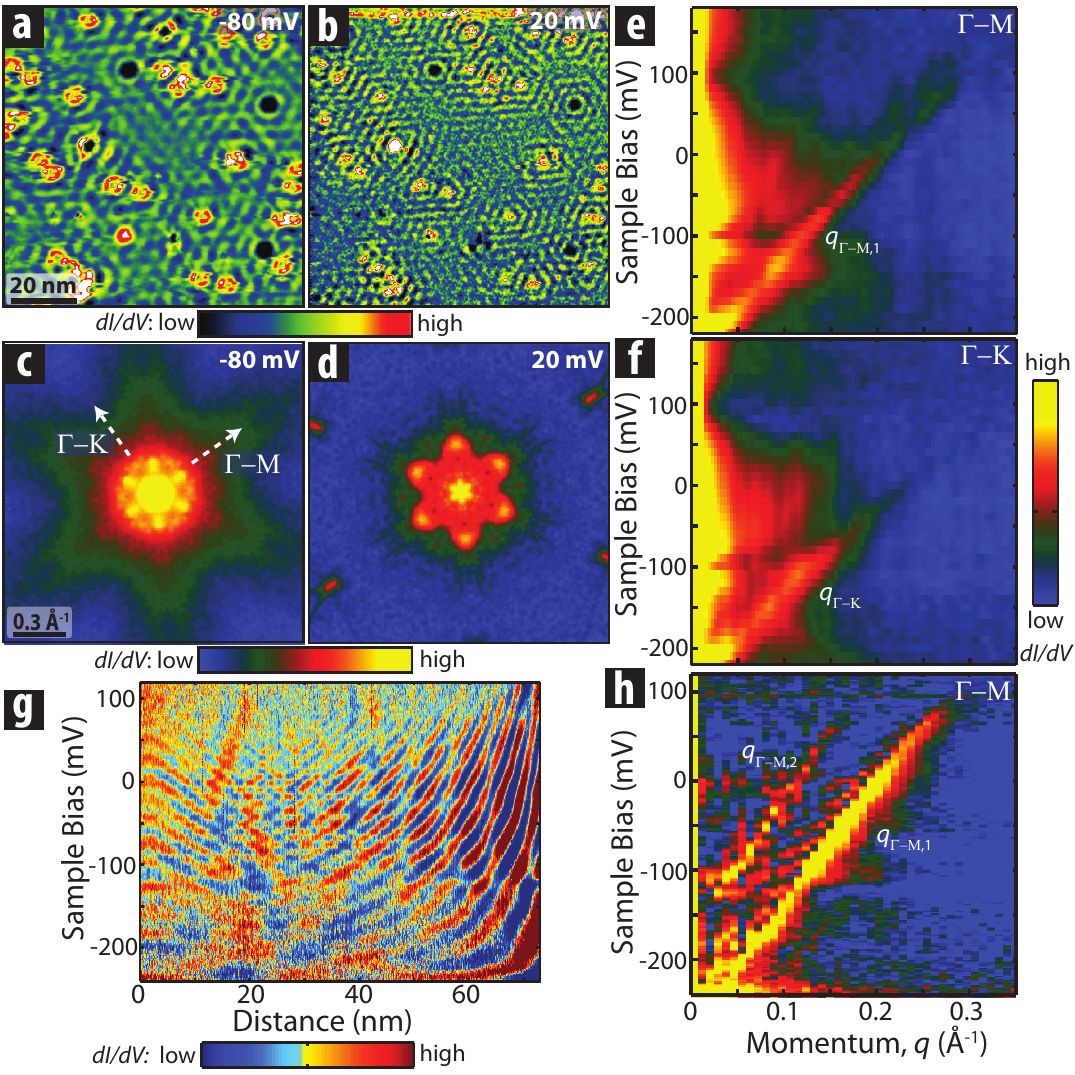}

\caption{\textbf{Quasiparticle Interference of Surface States.} \textbf{(a,
b)} $dI/dV(\vec{r},V)$ maps at sample bias -80~mV (a) and +20~mV
(b), exemplifying standing wave patterns generated by impurities.
\textbf{(c, d)} Fourier Transforms (FTs) of (a) and (b), showing distinct
conductance peaks along the $\Gamma-{\rm M}$ and $\Gamma-{\rm K}$
reciprocal directions. FTs have been six-fold symmetrized to improve
signal quality (Supp Info III). \textbf{(e, f)} Conductance linecuts
through the FTs along the $\Gamma-{\rm M}$ (e) and $\Gamma-{\rm K}$
(f) directions, generated from 190~nm spatial maps. The prominent
dispersing modes along each direction are labeled $q_{\Gamma-{\rm M},1}$
and $q_{\Gamma-{\rm K}}$. \textbf{(g)} Conductance linecut ($dI/dV(x,V)$)
perpendicular to an atomically sharp step, showing dispersing step
edge scattering (Supp Info IV). \textbf{(h)} FT of the conductance
in (g), showing two prominent dispersing modes along the $\Gamma-{\rm M}$
direction, labeled $q_{\Gamma-{\rm M},1}$ and $q_{\Gamma-{\rm M},2}$.\label{fig:qpi}}
\end{figure*}

\textcolor{black}{}%
\textcolor{black}{\ref{fig:qpi} shows the simultaneous observation
of QPI over a 300~meV energy range, which allows quantitative reconciliation
with LL spectroscopy, and complete reconstruction of the SS band structure
of Sb both above and below $\varepsilon_{{\rm S}}$. }The scattering
of SS quasiparticles from single-atom impurities on Sb(111) creates
interference patterns in $dI/dV(\vec{r})$ maps, exemplified in \ref{fig:qpi}a-b.
Fourier transformations of these patterns reveal prominent modes along
the $\Gamma-{\rm M}$ and $\Gamma-{\rm K}$ reciprocal directions
(\ref{fig:qpi}c-d) that disperse roughly linearly with energy over
$\sim350$~mV from the Dirac point (\ref{fig:qpi}e-f). \ref{fig:qpi}f,
showing the $\Gamma-{\rm K}$ dispersion, extends previous reports
of the $\Gamma-{\rm M}$ dispersions\cite{Gomes2009,Seo2010,Yao2013}.
The scattering of SSs from atomic step edges creates similar interference
patterns (\ref{fig:qpi}g, Supp. Info. IV), allowing the extraction
of an additional dispersion along the $\Gamma-{\rm M}$ direction
(\ref{fig:qpi}h)\cite{Seo2010,Yao2013}. In the presence of a magnetic
field, no change is observed in the QPI. In particular, our measurements
over the same spatial region at magnetic fields of 0~T and 9~T show
no additional modes corresponding to LLs\cite{Biswas2009b} or field-induced
backscattering (Supp. Info. III), indicating the small magnitude of
the SS g-factor.

\begin{figure*}
\includegraphics[width=5.25in]{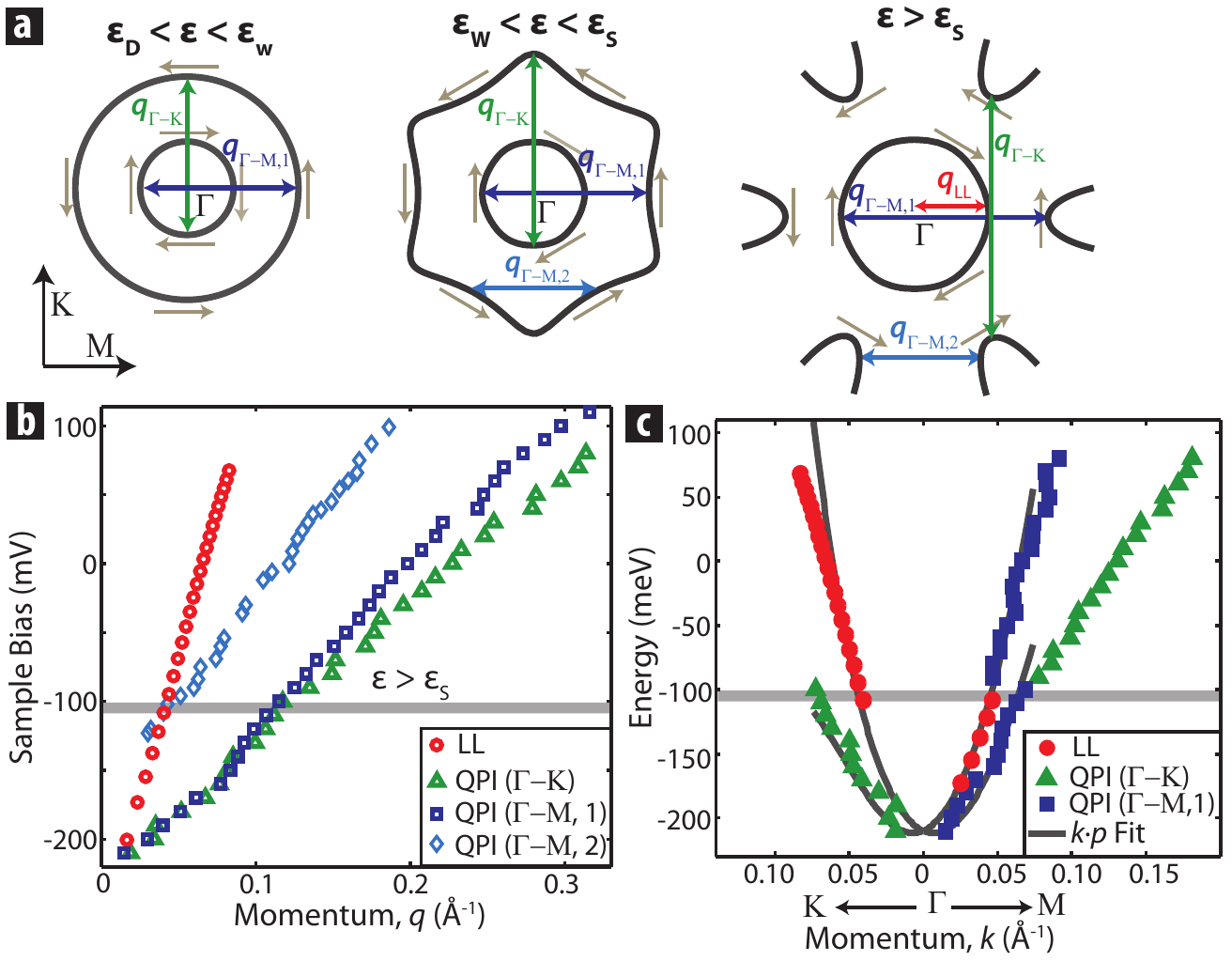}

\caption{\textbf{BSTM Band Structure.} \textbf{(a)} Schematic evolution of
the surface state contours of constant energy (CCEs), with the in-plane
spin polarization (brown), and $q$-space location of the dispersing
modes from LL \& QPI (detailed in (b)) overlaid. From the Dirac point
($\varepsilon_{{\rm D}}$) up to an energy $\varepsilon_{{\rm S}}$,
the CCEs correspond to a Rashba-split double 'cone\textquoteright{}
structure. The outer \textquoteleft{}cone\textquoteright{} acquires
a warped snowflake shape above an intermediate energy $\varepsilon_{{\rm W}}$.
Above $\varepsilon_{{\rm S}}$, the CCE topology changes, and the
outer SS band is no longer a closed contour. \textbf{(b)} A compilation
of three $\varepsilon(q)$ dispersions recorded over the same atomically
flat spatial region ($q_{\Gamma-{\rm M},2}$ is acquired from a nearby
step edge) using Landau quantization (\ref{fig:lls}b, red) and QPI
(\ref{fig:qpi}e-h, blue, green and cyan). \textbf{(c)} The BSTM dispersion
$\varepsilon(k)$ of the SS band structure, deduced from (a) and (b).
Grey lines correspond to a fit to the data using the $k\cdot p$ model
in \ref{eq:KPHamiltonian}\cite{Fu2009}\textcolor{black}{. \label{fig:stm_bands}}}
\end{figure*}

The $k$-space origin of the QPI modes are indicated on the schematic
CCEs in \ref{fig:stm_bands}a, which display three qualitatively different
shapes over the energy range of interest. Their $q(\varepsilon)$
dispersions are plotted in \ref{fig:stm_bands}b. First, across the
entire observed energy range, the dominant $\Gamma-{\rm M}$ QPI mode,
$\vec{q}_{\Gamma-{\rm M},1}(\varepsilon)$, corresponds to inter-\textcolor{black}{band}
scattering between parallel spins. Meanwhile, the second $\Gamma-{\rm M}$
mode, $\vec{q}_{\Gamma-{\rm M},2}(\varepsilon)$, involves intra-band
scattering across the outer band, which comes into play only for $\varepsilon>\varepsilon_{{\rm {\rm w}}}$,
where it is sufficiently warped\cite{Seo2010}. Finally, the $\Gamma-{\rm K}$
QPI mode $\vec{q}_{\Gamma-{\rm K}}(\varepsilon)$ corresponds to inter-band
scattering for $\varepsilon_{{\rm B}}<\varepsilon<\varepsilon_{{\rm {\rm S}}}$,
where both CCEs are nearly circular, and is therefore identical to
$\vec{q}_{\Gamma-{\rm M},1}(\varepsilon)$ at these energies. The
kink in $\vec{q}_{\Gamma-{\rm K}}(\varepsilon)$ around $\varepsilon_{S}$
corresponds to a crossover to scattering between the 'pocket'-like
sections of the outer band for $\varepsilon>\varepsilon_{S}$.

\begin{figure*}
\includegraphics[width=4.25in]{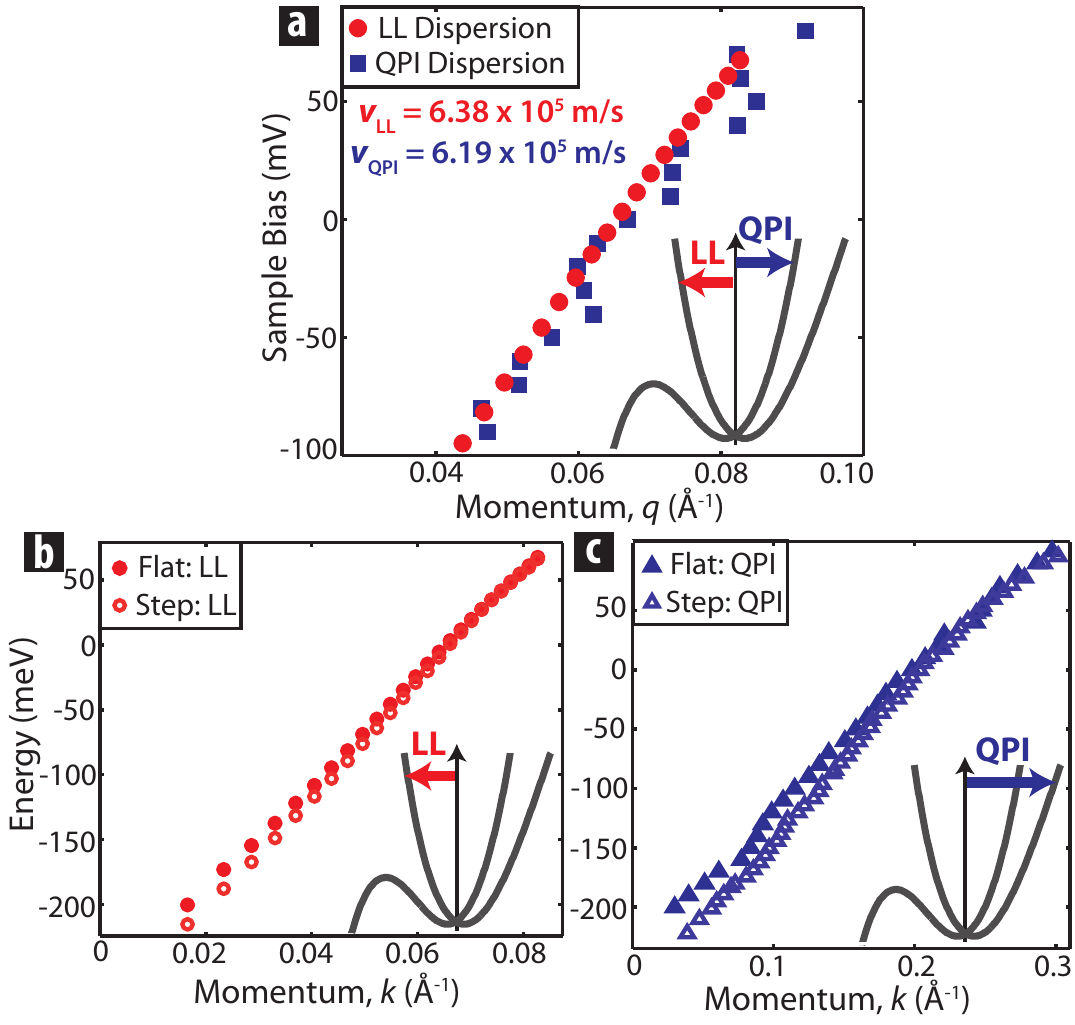}

\caption{\textbf{BSTM: Consistency and Spatial Resolution.} \textbf{(a)} The
dispersion of the inner SS cone, measured using LL and QPI techniques
over the same spatial region. The sign of the observed $\sim3\%$
difference between the techniques is consistent with hexagonal warping,
which results in a difference in dispersion along the $\Gamma-{\rm M}$
and $\Gamma-{\rm K}$ directions. The inset shows schematic SS bands
in grey, with the horizontal arrows indicating the measured $k(\varepsilon)$
for each panel. \textbf{(b-c)} Nanoscale spatial sensitivity of BSTM
demon\textcolor{black}{strated by comparing the (b) dispersion of
the inner band deduced from LLs and (c) Dispersion of the outer band
deduced from QPI ($q_{\Gamma-M,1}$) and LLs. Dispersions were recorded
over atomically flat regions (filled points) and terraced regions
(hollow points), $\sim200$~nm away from each other. Both (b) and
(c) indicate a }\textbf{\textcolor{black}{\emph{consistent, non-rigid
difference}}}\textcolor{black}{{} in the band structure between the
two regions, demonstrated by the offset and slope change between the
two curves.} \label{fig:llqpi_spatial}}
\end{figure*}

\textcolor{black}{}%
\textcolor{black}{{} }Having determined the origin of all observed $q$-vectors,
we use \textcolor{black}{$\vec{q}_{\Gamma-{\rm K}}(\varepsilon)$
and $\vec{q}_{\Gamma-{\rm M},1}(\varepsilon)$} to extract the dispersions
of both cones for \textcolor{black}{$\varepsilon>\varepsilon_{{\rm S}}$}
\textcolor{black}{(\ref{fig:stm_bands}c, details in Supp. Info. V)},
for direct comparison with the inner cone dispersion extracted from
Dirac LLs over the same 200 mV energy range (\ref{fig:lls}b). \textcolor{black}{The
independently measured inner cone dispersions are consistent to within
3\% (\ref{fig:llqpi_spatial}a), the sign of the deviation being in
agreement with that expected from hexagonal warping effects\cite{Fu2009}.
Furthermore, both LL and QPI measurements agree with ARPES measurements
of filled state dispersion to within 10\%, comparable to the variation
between independent ARPES measurements\cite{Hsieh2010,Gomes2009}.
We thus reconcile the techniques of LL spectroscopy and QPI imaging,
and establish their quantitative credibility as momentum space probes. }

\textcolor{black}{}%
\textcolor{black}{{} Upon resolving this discrepancy which has limited
the combined use of LL and QPI techniques, we proceed to use them
in concert, exploiting their complementary sensitivity to different
$k\cdot p$ parameters to determine the band structure in the Rashba
($\varepsilon<\varepsilon_{{\rm S}}$) regime (\ref{fig:stm_bands}c).
We find that our $\varepsilon<\varepsilon_{{\rm S}}$ data is best
reproduced in the $k\cdot p$ description with: $\varepsilon_{{\rm D}}=-210$~meV,
$m^{*}=0.1\, m_{e}$, $\alpha=110$~Å$^2$, $\lambda=230$~eV·Å$^3$,
and the crucial spin-orbit coupling, $v_{0}=0.51$~eV·Å ($7.7\times10^{4}$~${\rm m/{\rm s}}$).}

\textcolor{black}{}%
\textcolor{black}{We thus present a proof-of-principle demonstration
establishing band structure tunneling microscopy (BSTM) -- a combination
of LL and QPI spectroscopy which is crucial to the nanoscale reconstruction
of multi-component band structures of 2D electronic materials. In
contrast to previous STM work\cite{Dombrowski1998}, our QPI patterns
extend far beyond individual scatterers (\ref{fig:qpi}a-b), and are
thus are independent of impurity models and compatible with the Friedel
approximation\cite{Friedel1958}. In contrast to ARPES, BSTM can probe
empty states -- without sacrificing energy resolution by populating those
states thermally. Crucially, we demonstrate the nanoscale spatial
sensitivity of BSTM by showing up to 5\% }\textbf{\textcolor{black}{\emph{non-rigid}}}\textcolor{black}{{}
band structure changes between atomically flat and terraced regions
separated by $\sim200$~nm (\ref{fig:llqpi_spatial}b-c}), after
ruling out tip-induced artifacts (Supp. Info. V). The step edges must
have broken bonds, which may cause charge redistribution as well as
structural distortion, either of which may bear responsibility for
these non-rigid spatial variations.

\section{Discussions\label{sec:discussions}}

Our establishment of BSTM on Sb(111) sheds light on several fundamental
and practical issues directing the exploration of topological materials.
First, the existence of up to 27 LLs arising from a single, robust
cone \textendash{} despite the presence of proximate surface and bulk
bands throughout the energy range \textendash{} is surprising. It
had been speculated that in the Bi$_{2}$X$_{3}$ class of topological
materials, the onset of bulk bands induces surface-bulk scattering,
limiting the observed range of LLs\cite{Hanaguri2010}. In contrast,
our demonstration of robust Landau quantization in a semimetal suggests
that even in the presence of proximate bulk bands, closed SS contours
exhibit a long lifetime, suggesting that that they maintain their
topological protection against \textcolor{black}{inelastic scattering,
in addition to backscattering. Second, the use of topological materials
for spintronics devices will require strong spin-momentum locking,
long mean free path $\ell_{{\rm f}}$, and small g-factor -- parameters
which can be quantified by BSTM. We} note the quantitative distinction
between the Rashba parameter ($v_{0}=0.51$~eV·Å, extracted from
the $k\cdot p$ fit) and the Dirac velocity $v_{{\rm D}}=4.2$~eV·Å
(\ref{fig:llqpi_spatial}a), and clarify that the former is the spin-momentum
locking parameter relevant towards spintronics applications. Meanwhile,
LL widths place a lower bound on $\ell_{{\rm f}}$, while the absence
of field-induced backscattering QPI channels places an upper bound
on $g$. Third, the search for better topological materials has gravitated
towards insulating ternary and quaternary materials, tuned off-stoichiometry
to enhance the SS contribution in transport measurements\cite{Arakane2012,Jia2012}.
However, our observations suggest that the presence of a bulk continuum
suppresses chemical potential fluctuations\cite{Jia2012}, actually
enhancing the mean free path rather than diminishing the lifetime
of SS quasiparticles as had previously been speculated\cite{Kim2011,Hanaguri2010}\textcolor{black}{.
We therefore suggest heterostructures, with the appropriate use of
semimetals, as an alternate avenue towards better topological devices
with immunity to disorder\cite{Yao2013}}\textcolor{blue}{.}

\noindent %
Our simultaneous spatial and spectral observation of LLs and QPI,
followed by their quantitative reconciliation, establish BSTM as a
reliable, self-consistent nanoscale band structure probe. Unique advantages\textcolor{black}{{}
of BSTM include nanoscale sensitivity to band structure deformations,
accessibility of empty states, and utility in magnetic field. We therefore
underscore the crucial role that BSTM can play in characterizing diverse
electronic compounds and growth techniques, as well as developing
nanoscale devices using heteroepitaxial van der Waals materials\cite{Novoselov2005b}.
In particular, we suggest that Sb, with its particularly long-lived
surface states, may be an excellent choice for investigating the spatial
evolution of topological proximity effects\cite{Hasan2010,Qi2011}.}

\section{Methods\label{sec:methods}}

\noindent \textbf{\emph{Sample Growth.}} Single crystals of Sb were
grown using the following method\cite{Hsieh2009b}. High-purity antimony
(99.999\%, supplied by Alfa Aesar\textregistered{}) in shot form (10.15~g,
6~mm) was sealed in an evacuated quartz tube, and heated in a box
furnace to 700~\textdegree{}C for 24~hours. The furnace was cooled
slowly (0.1~\textdegree{}C/min) to 500~\textdegree{}C, and subsequently
cooled to room temperature.

\noindent

\noindent \textbf{\emph{STM Measurements.}}\textbf{ }Our measurements
were performed using a home-built STM at liquid helium temperatures.
Single crystals of Sb were cleaved in-situ in cryogenic ultrahigh-vacuum
to expose the (111) face, and inserted into the STM. Mechanically
cut Pt-Ir tips, cleaned by field emission and characterized on gold,
were used for the measurements. Spectroscopy data were acquired using
a lock-in technique at 1.115~kHz, and conductance maps were obtained
by recording out-of-feedback $dI/dV$ spectra at each spatial location.
Three samples were investigated in this work, and their correspondence
to the data shown in the manuscript is detailed in Supp. Info. I.

\paragraph*{\noindent \textbf{Acknowledgements.}}

\noindent We are grateful to Anton Akhmerov, Liang Fu, Bert Halperin,
Vidya Madhavan, Joel Moore, and Jay Sau for insightful discussions.
The work at Harvard was supported by the NSF under Grant No. DMR-1106023,
and by the New York Community Trust - George Merck Fund. The work
at MIT was supported by US Department of Energy, Office of Science,
Office of Basic Energy Sciences under Grant No. DE-FG02-07ER46134.
In addition, we acknowledge funding from A{*}STAR, Singapore (A.S.);
NSERC, Canada (M.M.Y.); and the Singapore NRF, under Award No. NRF-NRFF2013-03
(H.L.).

\paragraph*{\noindent \textbf{Author Contributions.}}

\noindent A.S., M.M.Y. and Y.H. performed STM experiments and A.S.
analyzed the data, supervised by J.E.H. D.R.G. grew the samples, supervised
by Y.S.L. H.L. performed calculations with advice from A.B. A.S. and
J.E.H. wrote the manuscript.

\rule[0.5ex]{0.8\columnwidth}{1pt}

\clearpage{}
\end{document}